\documentclass[fleqn,10pt]{wlscirep}
\usepackage[utf8]{inputenc}
\usepackage[T1]{fontenc}
\title{Sm-Nd Isotope Data Compilation from Geoscientific Literature Using an Automated Tabular Extraction Method}

\author[1]{Zhixin Guo}
\author[2]{Tao Wang}
\author[2]{Chaoyang Wang}
\author[1]{Jianping Zhou}
\author[3,*]{Guanjie Zheng}
\author[1]{Xinbing Wang}
\author[4]{Chenghu Zhou}
\affil[1]{Shanghai Jiao Tong University, School of Electronic Information and Electrical Engineering, Shanghai, 200240, China}
\affil[2]{Chinese Academy of Geological Sciences, Institute of Geology, Beijing, 100037, China}
\affil[3]{Shanghai Jiao Tong University, John Hopcroft Center for Computer Science, Shanghai, 200240, China}
\affil[4]{Chinese Academy of Sciences, Institute of Geological Sciences and Natural Resources Research, Beijing, 100101, China}

\affil[*]{corresponding author(s): Guanjie Zheng (gjzheng@sjtu.edu.cn)}

\begin{abstract}
The rare earth elements Sm and Nd significantly address fundamental questions about crustal growth, such as its spatiotemporal evolution and the interplay between orogenesis and crustal accretion. Their relative immobility during high-grade metamorphism makes the Sm-Nd isotopic system crucial for inferring crustal formation times. Historically, data have been disseminated sporadically in the scientific literature due to complicated and costly sampling procedures, resulting in a fragmented knowledge base. However, the scattering of critical geoscience data across multiple publications poses significant challenges regarding human capital and time. In response, we present an automated tabular extraction method for harvesting tabular geoscience data. We collect 10,624 Sm-Nd data entries from 9,138 tables in over 20,000 geoscience publications using this method. We manually selected 2,118 data points from it to supplement our previously constructed global Sm-Nd dataset, increasing its sample count by over 20\%. Our automatic data collection methodology enhances the efficiency of data acquisition processes spanning various scientific domains. Furthermore, the constructed Sm-Nd isotopic dataset should motivate the research of classifying global orogenic belts.
\end{abstract}
\begin{document}

\flushbottom
\maketitle

% \thispagestyle{empty}

% \noindent Please note: Abbreviations should be introduced at the first mention in the main text – no abbreviations lists or tables should be included. Structure of the main text is provided below.

\section{Background \& Summary}
In igneous science, it is critical to efficiently collect spatial, temporal, and geochemical data from myriad samples. Such data play a prominent role in understanding high-level geological phenomena such as reconstructing ancient plate kinematics, delineating processes of continental convergence and dispersal, assessing crustal growth, and investigating the deep material structure of orogens \cite{tarney1994trace,condie2014growth,cawood2009accretionary,carminati2012geodynamic,wang2023quantifying, wang2023quantitative, wang2023voluminous}. Determining the formation time of continental crustal protoliths within metamorphic terrains is a significant challenge. This complexity is primarily due to the fluctuating effects of tectonothermal events on isotopic systems. Such events can lead to redistributing parent-daughter elements and consequently to partial or complete re-equilibration of isotopic systems on scales ranging from individual minerals to entire outcrops. In particular, Sm and Nd are often considered relatively immobile during high-grade metamorphism \cite{green1969rare, pride1980rare}. As a result, the Sm-Nd isotopic system has been used extensively to derive ages ranging from the most recent metamorphic events to the original formation of the crust \cite{wasserburg1980early, moorbath1975isotopic, hamilton1979sm, jacobsen1978interpretation}. Groundbreaking studies have highlighted the importance of the Sm-Nd isotope. These studies have investigated the provenance and petrogenesis of igneous rocks \cite{gruau1991origin,lambert1994re}, identified allochthonous terranes within orogens \cite{dickin1998nd,dickin2000crustal, wang2009nd}, and investigated the dynamics of continental crustal growth \cite{goldstein1984sm,blanchet2019database,condie2013preservation,cawood2009accretionary,cawood2016linking,collins2011two,collins2002hot}. Compilations of Sm-Nd isotope datasets have proven to be invaluable repositories, providing insights into historical and contemporary states of global systems and offering predictions of future conditions. Orogens, the foundations of plate tectonics, are primarily composite structures that undergo myriad phases of orogeny. 

Despite its potential, synthesizing Sm-Nd isotope data, which spans vast spatial and temporal dimensions and cuts across diverse scientific landscapes, presents formidable challenges. These arise primarily from the nuanced processes required to extract large datasets from various scientific publications efficiently. By their very nature, geoscience data are besieged by a number of obstacles, from the pitfalls of non-reproducibility and inherent uncertainties to the challenges posed by their multifaceted nature, massive computational demands, and the need for cyclical updates \cite{sudmanns2020big}. In the burgeoning era of big data, the international geoscience community is grappling with an unprecedented surge in the volume and complexity of data structures \cite{cai2015challenges, ahmed2017role}. Traditional methods, which rely primarily on manual data collection and categorization, need to be revised when confronted with the colossal repositories of information and the multifaceted challenges of data archival review \cite{miller2009geographic,chen2011design,last2002tracking} Despite the increased emphasis that modern magmatic databases place on both the volume and integrity of their samples, the effectiveness of data collection strategies remains underexplored \cite{sarbas2008georoc,gard2019global}. 

With the advent of neural network-based techniques, there has been a marked alleviation in the challenges associated with analyzing vast textual data. A significant trend has emerged wherein there is an inclination toward deciphering elements and extracting semantic content from PDF documents. The techniques propelling these advancements can be bifurcated into two primary domains: methods anchored in visual analysis \cite{ren2015faster,he2017mask,redmon2018yolov3} and those founded on pre-trained language models (PLMs) \cite{garncarek2021lambert,yu2021pick,xu2020layoutlm,zhang2020trie,rahman2020integrating}. Both approaches have showcased noteworthy efficacy in their designated domains. Notwithstanding, it is pivotal to note that these techniques predominantly concentrate on visual element recognition or deeply engage with natural language processing (NLP) in end-to-end networks. Although proficient within their respective scopes, such specialization may not fully cater to the burgeoning demand for swift and exhaustive data extraction from intricate tabular formats. Historically, methodologies, as exemplified by platforms like Chronos \cite{cervato2005chronos}, GeoSciNet \cite{snyder2008geoscinet}, GeoDeepDive \cite{zhang2013geodeepdive}, SciSpace \cite{khan2019scispace}, and GeoDeepShovel \cite{zhang2022geodeepshovel}, have leaned on neural network paradigms explicitly tailored for scientific literature analysis. However, despite their significant contributions, most of these methods are predominantly tailored for single-document analysis, consequently constraining their applicability to data extraction from individual documents.

In light of this limitation, we present an innovative approach, illustrated in Figure~\ref{workflow}, that bifurcates the process into two core modules: document retrieval and tabular data aggregation. Unlike these efforts in geoscience data aggregation \cite{niu2014ontology, walker2005earthchem, boone2022ausgeochem, rodriguez2022colombian}, our approach provides a more cohesive workflow that seamlessly integrates document retrieval and tabular data collection. Such integration dramatically increases the efficiency of data acquisition from geoscience repositories. Using this tool, we obtained 10,624 Sm-Nd datasets from 9,138 tables spanning more than 20,000 geoscience articles. Extensive curation of this vast repository resulted in 2,118 meticulously curated datasets, adding over 20\% to our global Sm-Nd compendium.

\section{Methods}
\subsection{Document Retrieval}

\begin{figure*}[!t]
\centering
\includegraphics[scale=0.52]{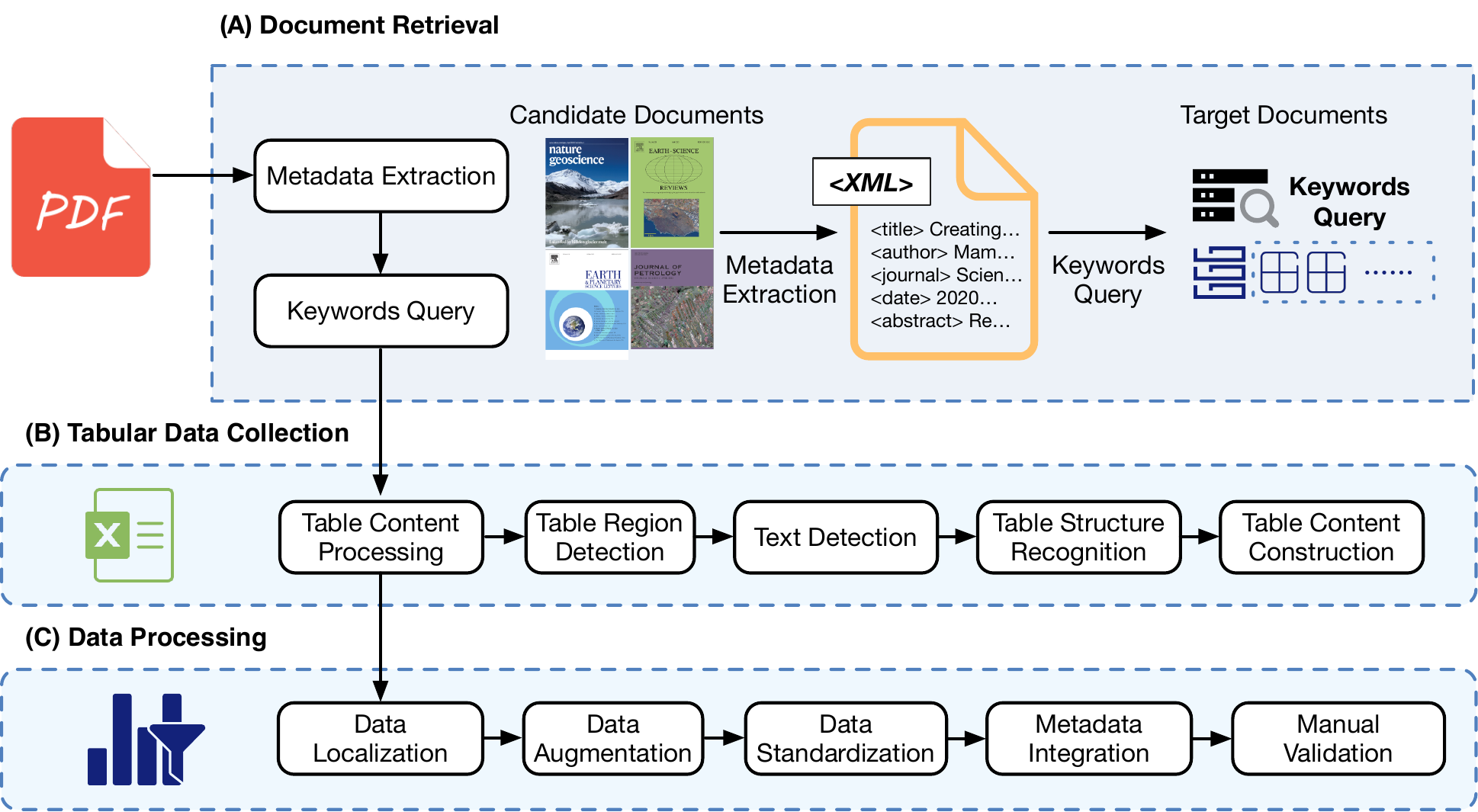}
\caption{An overview of the automatic searching and collecting tabular data tool workflow, which consists of two main steps: (A) document retrieval and (B) tabular data collection.}
\label{workflow}
\end{figure*}

Figure~\ref{workflow}(A) illustrates the core components of our PDF parsing protocol, which integrates metadata extraction and advanced keyword querying techniques. The beginning of our process is anchored in the meticulous extraction of metadata from a curated selection of PDF documents. In this phase, we use CERMINE \cite{tkaczyk2015cermine}, an automatic tool equipped with precision to extract structured metadata from various scientific publications. This instrumental phase ensures a comprehensive and accurate capture of key metadata components, including but not limited to title, authorship, abstract, affiliations, included keywords, and intricate bibliographic specifics (including journal name, volume, issue, and DOI). A notable feature of CERMINE is its ability to capture the rich content structure, encapsulating textual narratives, segmented sections, illustrative figures, tabular data, and accompanying captions while preserving the integrity of layout specifics and element positions. The captured metadata is then transformed into XML documents to facilitate an enhanced scheme of analytical review and organizational refinement.

The following phase is characterized by optimizing the keyword query efficiency, which is realized by implementing a multifaceted data transformation protocol. Integral to this process is the conversion of all tokens to lowercase, the substitution of spaces for non-alphanumeric characters, the elimination of common stop words, and the expansion of standard abbreviations, culminating in a standardized and normalized data set that enhances consistency and comparative accuracy in subsequent phases of data retrieval. In the context of our focus study on Sm-Nd isotopes, a meticulously constructed keyword matrix, described in Table~\ref{keywords}, coupled with specialized filtering rules, enhances the precision and speed of target document identification and extraction.

\begin{enumerate}
    \item Inclusion criteria include terms such as $\epsilon$Nd, Sm-Nd, Sm-Nd-Hf, 143Nd/144Nd, Nd isotope, or TDM in the title or abstract.
    \item The criteria further extend to the inclusion of terms related to geological and mineralogical descriptors, including felsic, granite, granitic, pluton, plutonic, magmatic, magma, diorite, or rhyolite, in the title or abstract.
\end{enumerate}

Our methodology culminates in selecting scholarly articles that meet the two enumerated criteria, ensuring a comprehensive and targeted inclusion of documents intrinsically aligned with our study's thematic and conceptual underpinnings. This multifaceted approach underscores a synthesis of methodological rigor and technological sophistication, embodying a blueprint for enhanced accuracy and efficiency in extracting and analyzing complex scientific data.

\begin{figure*}[!t]
\centering
\includegraphics[scale=0.4]{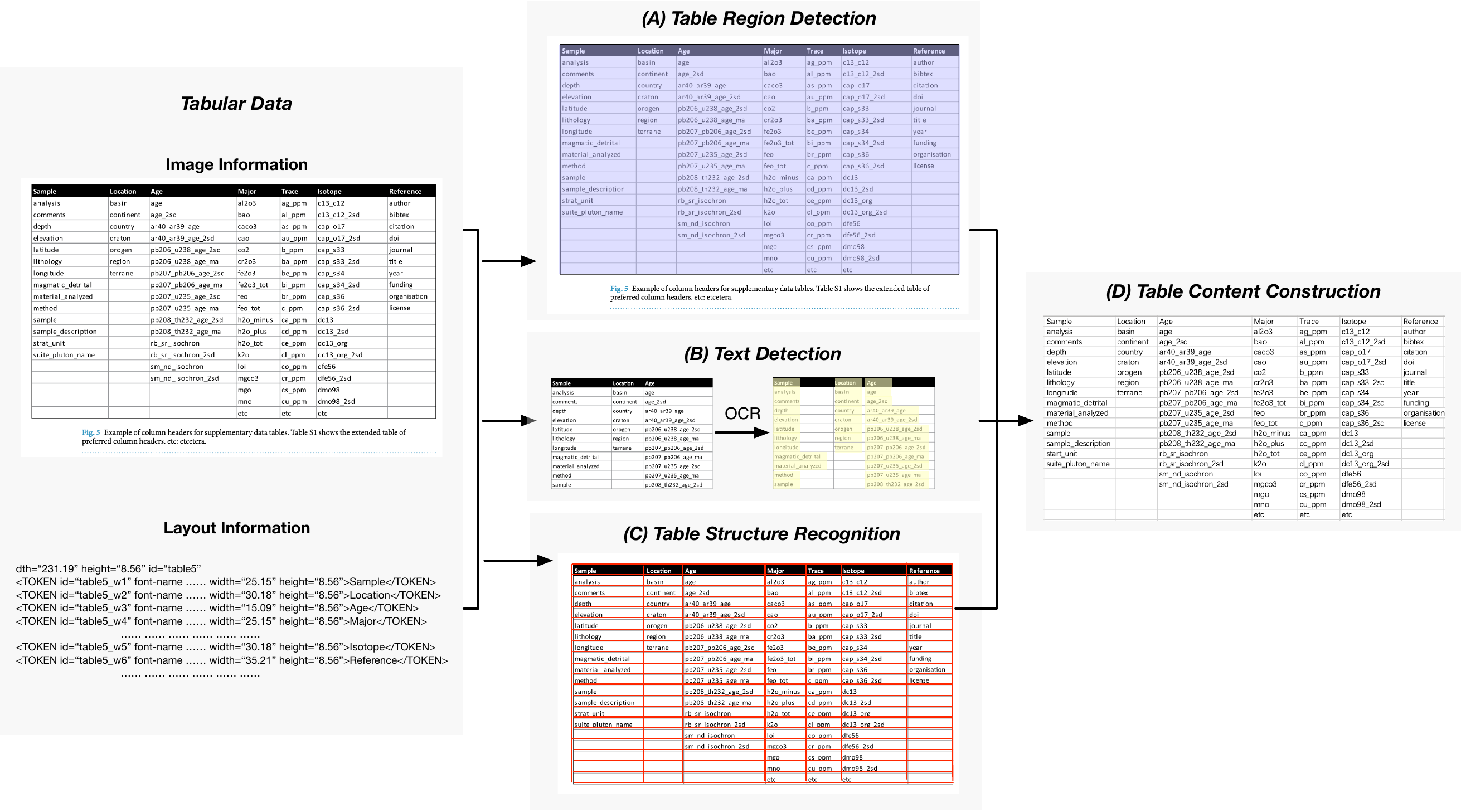}   
\caption{An overview of tabular data collection pipeline. The pipeline process consists of four modules: (a) table region detection, (b) text detection, (c) table structure recognition, and (d) table content construction.}
\label{tableExtraction}
\end{figure*}

\subsection{Tabular Data Collection}
Figure~\ref{tableExtraction} illustrates the pipeline we developed for extracting tabular data from PDF documents, a complex task due to the heterogeneous nature of these files. The elaborated strategy is based on three key stages: table region detection, table structure recognition, and nuanced table content construction. A notable challenge in this endeavor is a universal encoding standard for PDF files. This gap leads to problems, particularly the appearance of garbled text, which leads to omitting essential layout information. We address this challenge by converting each page of tabular data to images. We then employ a computer vision (CV)-based methodology to increase the recognition accuracy for table construction.

\begin{table}[htbp]
\begin{center}
\begin{tabular}{|l|l|l|l|l|}
\hline
Sample        & Nation           & Region        & GeoTectonic unit & Groups    \\ \hline
Tectonic unit & Subtectonic unit & Sub groups    & Longitude        & Latitude  \\ \hline
Lithology     & Pluton           & Formation     & Age (Ma)         & Sm        \\ \hline
Nd            & 147Sm/144Nd    & 143Nd/144Nd & $2\sigma$               & $f_{Sm/Nd}$    \\ \hline
$\epsilon$Nd       & TDM1             & TDM2          & Author      & Year \\ \hline
Journal  & Title            & Volume        & Page             & DOI       \\ \hline
\end{tabular}
\end{center}
\caption{Keyword list for document identification and extraction.}\protect
\label{keywords}
\end{table}

\subsubsection{Table Region Detection}
Our table region detection model was developed using Fast Region-based Convolutional Network (Fast R-CNN) \cite{girshick2015fast}, a machine learning algorithm acclaimed for its accuracy and efficiency. Fast R-CNN is instrumental in accurately identifying and precisely localizing table regions in various documents. The model's effectiveness is further enhanced by its training on TableBank \cite{li2020tablebank}, a robust and comprehensive dataset curated explicitly to refine table detection algorithms. Our model demonstrated 97\% accuracy on the TableBank dataset throughout the training phase. 

\begin{figure*}[!t]
\centering
\includegraphics[scale=0.32]{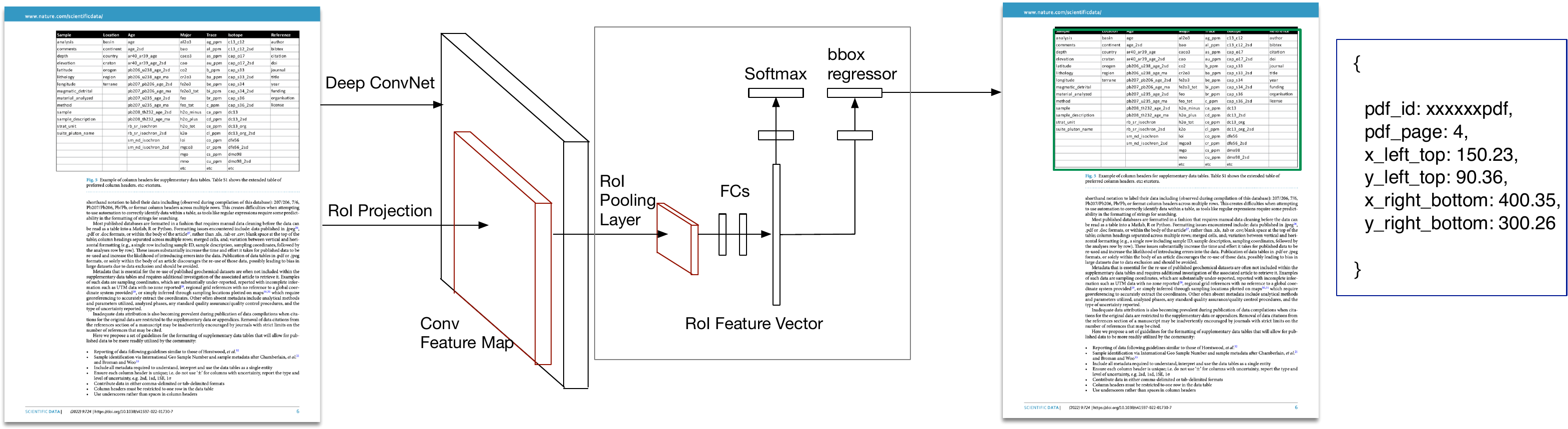}   
\caption{An overview of table region detection processing. The table detection neural network localize tables and returns the table position and page information.}
\label{tabledetection}
\end{figure*}

As a practical application of the refined model, we set out to identify tables embedded in academic PDFs. Using PDF parsing techniques, we extract metadata to detect the presence of embedded table tags within document pages. When such tags are detected, the corresponding pages are converted to images, setting the stage for more granular analysis. Figure~\ref{tabledetection} presents an overview of table region detection processing. The Table Detection Neural Network then analyzes each image to identify and localize tables accurately. In cases where the network confirms the presence of a table, it not only returns the specific page number but also pinpoints the exact relative position of the table frame on that page. 

\subsubsection{Text Detection}
Our methodology begins with an assessment to determine if the PDF pages with tables have an accompanying text layer. In cases where this layer is missing, we employ easyOCR \cite{jeeva2022intelligent}, a reputable Optical Character Recognition (OCR) tool celebrated for its precision and efficiency, to introduce a text layer. This addition ensures that the text aligns accurately with its corresponding positions in the image. Once integrated, we meticulously organize the text layer to maintain visual and structural integrity. Subsequently, an exhaustive inventory of text dimensions across the entire PDF document is prepared, offering an in-depth analysis of text characteristics.

The average text width, derived from this inventory, is instrumental in estimating text spacing - a critical parameter in our study. This estimated text spacing is not merely a numerical value but serves as a cornerstone for deciphering the intricate structure of embedded tables within the document. It plays a pivotal role in enhancing the accuracy and efficiency of table extraction and analysis, thus contributing profoundly to the rigor and precision of our research outcomes.

\begin{figure*}[!t]
\centering
\includegraphics[scale=0.28]{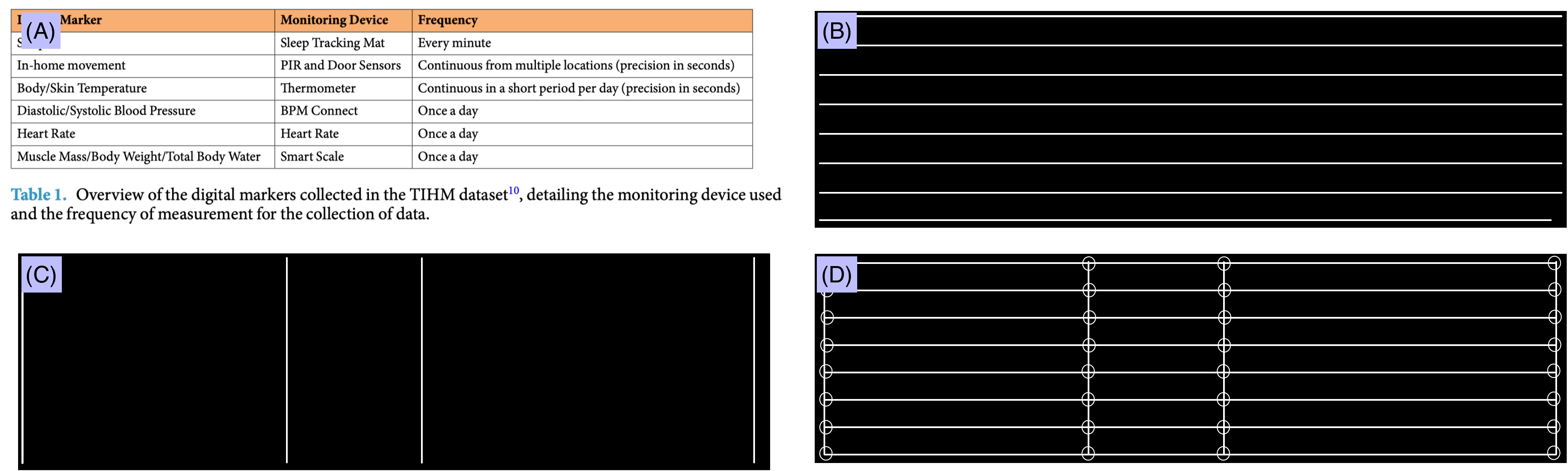}   
\caption{An overview of tabular structure recognition with frames. (A) denotes the initial table structure complete with frames. (B) indicates the detection of horizontal lines. (C) indicates the vertical line detection, and (D) showcases the identification of adjacent points.}
\label{withframe}
\end{figure*}

\subsubsection{Table Structure Recognition}
To accurately delineate sections within tables, our first approach is to capture images of tables, focusing on their outer borders and relative position within a PDF document. The images are then subjected to a sophisticated pre-processing stage. Each input image is first converted to grayscale, followed by adaptive thresholding for binarization. Morphological operations, namely image erosion and dilation, are then applied to identify vertical and horizontal lines. This meticulous process separates text and boundary lines while eliminating unnecessary noise. During the post-processing phase, we categorize table morphologies to determine the presence or absence of internal borderlines. This determination depends on detecting vertical and horizontal lines within the table. A pixel count that exceeds a predetermined threshold indicates the presence of these internal border lines. As shown in Figure~\ref{withframe}, we assiduously extract both vertical and horizontal lines, noting intersection points where they exist. Any superfluous intersections are methodically pruned, resulting in a streamlined set of internal frame intersection coordinates. We assess whether adjacent points connect to form internal frame lines using this refined dataset. These connections are established during validation, culminating in a structural representation that highlights tables equipped with internal frame lines.

\begin{figure*}[!t]
\centering
\includegraphics[scale=0.27]{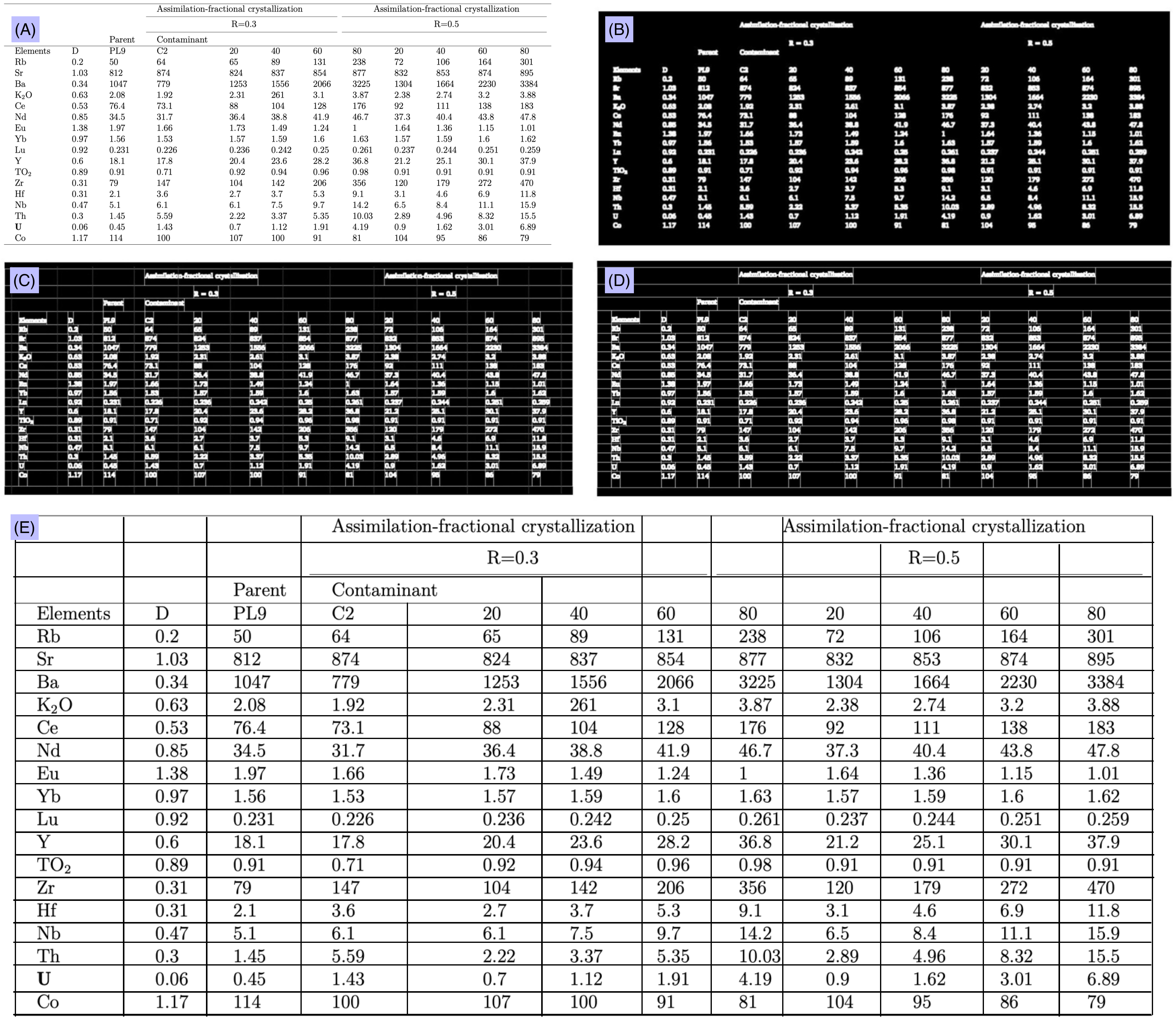}   
\caption{An overview of tabular structure recognition without frames. (A) presents the initial table lacking frames. (B) illustrates the result of image processing. (C) exhibits the construction of vertical and horizontal lines. (D) indicates the table structure post-cell merging recognition. (E) displays the final output following comprehensive table structure recognition.}
\label{noframe}
\end{figure*}

On the other hand, as shown in Figure~\ref{noframe}, for tables that do not have internal frame lines, our methodology focuses on their identification by analyzing the maximum connected intervals between texts. The procedure includes the following steps, ensuring a comprehensive approach to the extraction of complex tabular structures:
\begin{itemize}
    \item \textbf{Image Processing}. The input image is first converted to grayscale in the preliminary image preprocessing stage. Adaptive thresholding is then applied to achieve binarization. Subsequent operations include applying morphological techniques, specifically image erosion and dilation, to identify vertical and horizontal lines. Lines exceeding a certain length, determined by a predefined threshold, are systematically eliminated. After these steps, the image is subjected to another thresholding process that normalizes the pixel values: the values are set to 0 for blank areas and 255 for areas populated with text. This process ensures optimal contrast and improved clarity within the image.
    \item \textbf{Horizontal Frame Lines Identification}. We use a systematic pixel-by-pixel scan from top to bottom of the image. Rows with a pixel value sum of zero are marked as potential zones containing horizontal internal frame lines. By merging all adjacent potential zones and locating the midpoints of these continuous zero-value regions, we derive the precise locations of the internal frame lines that segment rows within the table.
    \item \textbf{Vertical Frame Lines Identification}. A sequential vertical scan is performed between adjacent horizontal internal frame lines. Each column's summation of pixel values facilitates the distinction between text and blank areas. We merge contiguous blank areas that are vertically aligned and eliminate those that are the height of a single line. The vertical lines that intersect the largest cumulative blank areas are designated as the vertical internal frame lines of the table, with each iteration marking the intersecting areas as traversed until all blank areas are accounted for.
    \item \textbf{Table Structure Recognition}. The intersection of the border and the vertical inner border lines delineates the primary unit cells of the table. Each cell's vertical lines are carefully examined to determine their intersection with text-containing zones. Segments of vertical lines that intersect text are excised, resulting in a horizontal merging of cells. This rigorous process reveals the table's refined internal frame line structure after cell merging.
\end{itemize}

This systematic methodology ensures accurate and efficient identification and extraction of table structures, especially those without explicit internal boundaries, by analyzing and processing the interconnectedness of text intervals and boundaries.

\subsubsection{Table Content Construction}
Using the reconstructed internal border structure of the table, we determine the rectangular frame coordinates for each cell within the PDF document. We then extract text information from coincident areas of the PDF text layer. After eliminating and adjusting for spaces, we determine the exact content of the table cell. By integrating the internal border lines and the corresponding cell contents, we construct an Excel spreadsheet, ensuring that the data associated with the merged cells is meticulously documented and preserved.

\subsection{Data Processing}
After extracting the tabular data, we embarked on a rigorous data processing regimen concentrating on the Sm-Nd isotope data. The procedural steps included the following methods:

\begin{itemize}
    \item \textbf{Data Localization}. The initial step is to locate the table that contains the Sm and Nd ratio, uniquely defined as the ${ }^{147} \mathrm{Sm} /{ }^{144} \mathrm{Nd}$ field. This preliminary stage allows the extraction of the Sm-Nd isotope data from the current literature, supplemented by its unique identifier, referred to as the Sample ID.
    \item \textbf{Data Augmentation}. It is worth noting that the extracted sample data often lacks several key fields, resulting in a remarkably sparse dataset. Based on the sample number derived from the previous step, additional searches are performed in other tables to enrich specific fields, such as the geographic coordinates of the sample. In cases where certain fields remain empty, the title and abstract of the literature are examined according to predefined protocols to retrieve and embed potentially correlating data.
    \item \textbf{Data Standardization}. Building on the initial data obtained, it was imperative to implement a data normalization process specifically for the key datasets. This adjustment not only supports easy data categorization and interpretation but also paves the way for intensive exploration of the assimilated data. Recalculated measures not only allow for a unified parameter-based data conversion but also enhance the robustness of the dataset, ensuring its consistency with the original literature data, thereby solidifying the credibility of the derived data.
    
    For the computation of Nd values, we adopted the chondritic standards, precisely $\left({ }^{143} \mathrm{Nd} /{ }^{144} \mathrm{Nd}\right)_{\mathrm{CHUR}}=0.512638$, $\left({ }^{147} \mathrm{Sm} /{ }^{144} \mathrm{Nd}\right)_{\mathrm{CHUR}}=0.1967159$ \cite{jacobsen1984sm}, $\left({ }^{143} \mathrm{Nd} /{ }^{144} \mathrm{Nd}\right)_{\mathrm{DM}}=0.51315$, and $\left({ }^{147} \mathrm{Sm} /{ }^{144} \mathrm{Nd}\right)_{\mathrm{DM}}=0.21372$, as representative of the contemporary Depleted mantle \cite{white1982sr,peucat1989sr}. 

    The analytical process focuses on the calculation of three primary standardized parameters, including $\varepsilon_{\mathrm{Nd}}$, $\mathrm{T}_{\mathrm{DM} 1}$, and $\mathrm{T}_{\mathrm{DM} 2}$. These are derived according to the following mathematical formulations \cite{depaolo1976nd}:
    
    \begin{equation}
    \varepsilon_{\mathrm{Nd}}=\frac{\left({ }^{143} \mathrm{Nd} /{ }^{144} \mathrm{Nd}\right)_{\text {sample }(\mathrm{T})}}{\left({ }^{143} \mathrm{Nd} /{ }^{144} \mathrm{Nd}\right)_{\mathrm{CHUR}(\mathrm{T})}-1} \times 10000
    \end{equation}

    \begin{equation}
    \mathrm{T}_{\mathrm{DM} 1}=1 / \lambda \times \operatorname{In}\left(1+\frac{\left({ }^{143} \mathrm{Nd} /{ }^{144} \mathrm{Nd}\right)_{\text {sample }(\mathrm{T})}-0.51315}{\left({ }^{147} \mathrm{Sm} /{ }^{144} \mathrm{Nd}\right)_{\text {sample }(\mathrm{T})}-0.2137}\right)
    \end{equation}

    \begin{equation}
    \mathrm{T}_{\mathrm{DM} 2}=\frac{\mathrm{T}_{\mathrm{DM} 1}-\left(\mathrm{T}_{\mathrm{DM} 1}-\mathrm{t}\right) \times\left(\mathrm{f}_{\mathrm{cc}}-\mathrm{f}_{\mathrm{s}}\right)}{\mathrm{f}_{\mathrm{cc}}-\mathrm{f}_{\mathrm{dm}}}
    \end{equation}

    The parameters \( f_{cc} \), \( f_s \), and \( f_{dm} \) represent the \( f_{Sm/Nd} \) values of the average continental crust, the sample, and the depleted mantle, respectively. Specifically, \( f_{cc} \) is set at 0.4, \( f_{dm} \) is 0.08592, and \( t \) signifies the intrusive age of the granite. The calculation for \( f_{Sm/Nd} \) is derived from the following equation:

    \begin{equation}
    \mathrm{f}_{\mathrm{Sm} / \mathrm{Nd}}=\frac{\left({ }^{147} \mathrm{Sm} /{ }^{144} \mathrm{Nd}\right)_{\text {sample }}}{\left({ }^{147} \mathrm{Sm} /{ }^{144} \mathrm{Nd}\right)_{\mathrm{CHUR}}}-1
    \end{equation}

    The two-stage model ages are derived by assuming that the protolith of granitic magmas has a \( f_{Sm/Nd} \) ratio characteristic of average continental crust. This approach is particularly favored for felsic igneous rocks to account for changes in \( f_{Sm/Nd} \) ratios that can be induced by geological processes such as partial melting, fractional crystallization, magma mixing, and hydrothermal alteration, among others. Our recalculated data are in close agreement with the original figures. In particular, certain literature sources provide only limited measurements, including values of \( \epsilon Nd(t) \) and \( T_{DM} \), which we have directly incorporated into our research \cite{keto1987nd}.

    \item \textbf{Metadata Integration}. Drawing a parallel between the acquired data and the metadata of the original article, we began a seamless integration process. This merge links the article's metadata, including author, year, journal, title, volume, page, and DOI, to the table data, laying the groundwork for subsequent manual review and ensuring an impeccable data lineage during the augmentation process.
    \item \textbf{Manual Validation}. As a final step, meticulous manual evaluations are performed to confirm the usability and integrity of the data.
\end{itemize}

\section{Data Records}
All collected data presented in this study are available on the Figshare repository: \url{https://doi.org/10.6084/m9.figshare.24054231.v2} \cite{guo2023smnd}. This data set consists of 12 files. Descriptions of these data records are as follows.

The "Sm-Nd data collection (Automated)" file provides a comprehensive overview of data systematically extracted from the geoscientific literature. Each dataset within this compilation is linked to its primary source through a detailed metadata framework. This framework includes several elements: "Ref. Author" (representing the author of the referenced paper), "Ref. Year" (representing the year of publication), "Ref. Journal" (identifying the publishing journal), "Title" (specifying the title of the referenced article), "Volume" (the specific volume of the publication), "Page" (specifying the exact page of data extraction), and "DOI" (the Digital Object Identifier, a tool that ensures unambiguous source referencing). In addition to this basic data, the automated process also collects specific target data: It collects sample identification ("sample"), geographic coordinates, and location descriptors ("Nation/Region/GeoTectonic Unit/Groups", "Tectonic Unit", "Sub-Tectonic Unit/Sub-Groups", "Longitude", "Latitude"), character attributes, "Latitude"), character attributes ("Lithology", "Pluton"), chronological details ("Age(Ma)"), and isotopic measurements ("Sm", "Nd", "147Sm/144Nd", "143Nd/144Nd", "2$\sigma$", "fSm/Nd", "$\epsilon$Nd(t)", "TDM1", "TDM2").

The "Sm-Nd data collection (Annotation)" file encompasses a detailed description of 2,119 manually annotated data points. These have been integrated to enhance our previously established global Sm-Nd dataset. Through meticulous data tracing, we have ensured the completion and validation of all fields. To promote uniformity, we revised the terminologies: "Nation/Region/GeoTectonic unit/Groups" has been standardized to "GeoTectonic unit", while "Subtectonic unit/Sub groups" has been recategorized as "Subtectonic unit". Following a comprehensive recalculation and normalization process, we introduced parameters such as "Calcul. $\epsilon$Nd(0), Recal.$\epsilon$Nd(t), Recalcu. TDM1 (Ga), and Recalcul. TDM2 (Ga)". Drawing upon the foundational historical data labeled "Orig. $\epsilon$Nd(0), Origin $\epsilon$Nd(t), Origi. TDM1 (Ma), and Origi. TDM2 (Ma)", we have transitioned the data nomenclature to "Mapped $\epsilon$Nd(t) and Mapped TDM(Ga)".

The file titled "Table S1a Nd data of Altaids (CAOB) 20230105-4713" consolidates an extensive dataset containing 4,713 entries, detailing the Nd isotopic data of the felsic and intermediate igneous rocks from the Altaids (CAOB). As delineated in prior research, the Altaids is alternately referred to as the Altaid tectonic collage, Central Asian Orogenic Supercollage, Central Asian Fold Belt, or more commonly, the southern Central Asian Orogenic Belt (CAOB). Geographically situated between the Siberian, Baltic (East European Cratons), and Tarim-North China cratons, the Altaids represent the most expansive accretionary orogen globally and are a paramount locus for Phanerozoic crustal growth \cite{csengor1993evolution, csengor2018tectonics, sengor1996turkic, csengor2022altaids, yakubchuk2017evolution, mossakovsky1994central}.

The file "Table S1b Nd data of Cordillera 20220419  1235 data OK new data 20220816" provides a comprehensive dataset comprising 1,235 entries that elucidate the Nd isotopic data of the felsic and intermediate igneous rocks from the North American Cordillera. Spanning over 10,000 km along North America's west coast, the North American Cordillera Orogen stands as a quintessential representation of the Cordilleran (ocean-continent subduction) orogenic system. Its significant expanse and geological features have been foundational in shaping various tectonic and geodynamic theories, making it a paradigmatic example of an accretionary orogen \cite{burchfiel1975nature,decelles2004late,decelles2009cyclicity,dickinson2004evolution,yonkee2015tectonic,fitz2018cretaceous,chapman2021north}.

The file "Table S1c Nd data of Newfoundland 20220816 418 data OK" presents an exhaustive dataset of 418 entries detailing the Nd isotopic data of the silicic and intermediate igneous rocks from the Newfoundland Appalachians. In North America, the Appalachian Orogen is recognized as a Paleozoic accretion-type orogen. Its formation can be attributed to the accretion and collision of numerous juvenile terranes and ancient blocks, culminating in the significant continent-continent collision between Laurentian and Gondwanan in the Late Permian \cite{williams1979appalachian, williams1988tectonic}.

The file "Table S1d Nd data of Lachlan  20220816   411data ok"  offers a thorough compilation of 411 entries elucidating the Nd isotopic data of the felsic and intermediate igneous rocks associated with the Lachlan orogen. Situated in Australia, the Lachlan Orogen forms an integral component of the Paleozoic Tasman orogenic system, which extends through eastern Australia and Gondwana. The Thomson Orogen geographically delimits this orogen to the north, the Delamerian Orogen to the west, and the New England Orogen to the east. Its genesis (spanning 450–340 Ma) is attributed to the closure of back-arc basin systems behind a persistent subduction zone symbolized by the New England Fold Belt. Further, its formation involved the accretion of submarine fans, accretionary complexes, extinct volcanic arcs, oceanic crust, and the Tasmanian microcontinent. Conceptually, the Lachlan Orogen is discerned as either a composite accretionary orogen or an extensional accretionary orogen \cite{foster2000evolution, glen2007early, foster2009palaeozoic, collins1998evaluation}.

The file "Table S1e Nd data of Tehtyan Tibet  20220821"  provides an exhaustive dataset consisting of 1,576 entries detailing the Nd isotopic values associated with the felsic and intermediate igneous rocks from Tethyan Tibet. This region, known as the Himalaya Orogen, constitutes a significant section of the Tethys. It stands as Earth's preeminent collisional orogen, characterized by its exceptionally thick crust, ranging from 60–85 km, and its remarkable elevation, exceeding 4,000 meters. The genesis of this orogen can be traced back to the India-Asia collision, which commenced between 65–55 Ma. This monumental collision event was preceded by the Mesozoic accretion and collision of various terranes, including but not limited to Songpan, north Qiangtang, south Qiangtang, and Lhasa \cite{yin2000geologic,tilmann2003seismic,van2012greater,xu2013orogen,xu2015crustal,xiao2015new,xiao2017anatomy,hou2015genetic}.

The file "Table S1f  Nd data of Caledonides  20220816  302 data  ok"  offers a comprehensive dataset, encompassing 302 entries that illuminate the Nd isotopic values of the felsic and intermediate igneous rocks from the Caledonides. The Caledonian Orogen, colloquially termed the Caledonides, is manifest in Western Europe, extending as a direct continuum of the Appalachian Mountain chain located in North America. Evidence of orogenic activity is discernible in the northern territories of the British Isles, Scandinavia, Svalbard, eastern Greenland, and specific segments of north-central Europe. The composition of this orogen is intricate, encapsulating numerous terranes rooted in Precambrian foundations—such as the ca. 2.7 Ga Scourian gneiss—and subsequent layering of late Precambrian to early Palaeozoic sediments and volcanic substrates \cite{cartwright1989evolution, frost1985caledonian, mckerrow2000caledonian, van1998cambrian, torsvik1996continental, oliver2001reconstruction}.

The file "Table S1g Nd data of Variscides 1299  20220816" presents an exhaustive dataset, comprising 1,299 entries, which elucidates the Nd isotopic values associated with the felsic and intermediate igneous rocks of the Variscides. The Variscan Orogen, also called the Hercynian Orogen or Variscides, is strategically situated to the south of the Caledonian Orogen, spanning regions of Europe and North Africa. This belt, arising from the Paleozoic-era collision between Gondwana and Laurussia, facilitated the subduction and accretion of various crustal blocks, culminating in the formation of the supercontinent Pangea. The Variscides are emblematic for their juxtaposition of disparate blocks of continental crust and can be interpreted as a continuous accretion of continental crust to Laurussia \cite{matte1991accretionary, kroner2013two, stampfli2013formation, laurent2017protracted}. As such, the structural and formation dynamics of the Variscides stand in stark contrast to oceanic accretionary orogens, such as the Altaids or CAOB.

The file "Table S1h  Nd data of Qinling-Dabie 20220816 480 data  OK" provides a comprehensive dataset, encompassing 480 entries, that delineates the Nd isotopic values about the felsic and intermediate igneous rocks of the Qinling-Dabie orogen. Situated across central China, the Qinling-Dabie Orogen is a pivotal orogenic structure in Asia. This orogen is composed of four discernible terranes or blocks. Arranged from north to south, these include the North China Block (NCB), the North Qinling Belt (NQB), the South Qinling Belt (SQB), and the South China Block (SCB). These terranes are demarcated by notable geological structures: the Luonan-Luanchuan fault zone and the Shangdan and Mianlue sutures. The genesis of the Shangdan suture can be traced back to the closure of the Shangdan Ocean, a Prototethys offshoot, between 500–420 Ma, while the Mianlue suture emerged due to the closure of the Mianlue Ocean, an offshoot of the Paleotethys, in the interval of 300–220 Ma. Geophysical investigations reveal a notable characteristic of this orogen: the absence of mountain roots and a standard crustal thickness approximating 40 km \cite{mattauer1985tectonics, meng1999timing, zhang2001qinling, dong2016tectonic, meng2000geologic}. 

The file "Table S2.  Nd Isotopic areas and areal percents of 8 orogens  20220821" provides a systematic representation of the isotopic domains and provinces, detailing both their spatial extents and relative areal percentages across the eight specified orogens.

\section{Technical Validation}
\subsection{Consistency Validation}
To maintain the consistency of Sm-Nd data across all orogenic belts, we selected 2,118 entries subjected to human evaluation, and of these, 1,780 entries were validated, resulting in a consistency rate of 84.04\%.

Our evaluation was based on three primary criteria:
\begin{itemize}
    \item We recalibrated the $\epsilon$Nd values and Nd mode ages employing consistent parameters and juxtaposed them with the values delineated in the original article of \cite{goldstein1984sm}. Data points were deemed valid if the recalibrated results were closely aligned or were analogous to the original values.
    \item For samples that incorporated geological information within the table, we executed spatial casts to authenticate the spatial positioning of the orogenic belt. The data was validated upon successful correlation with the location details specified in the article.
    \item Given that the samples discussed in this study predominantly comprise medium-acidic rocks, it is noteworthy that the $\epsilon$Nd(t) values of the 2,118 felsic and intermediate magmatic rocks exhibit a robust linear association with TDM2. This underscores the efficacy of automatic data collection, suggesting its comparability to manually curated data, especially in isotope mapping studies.
\end{itemize}

\subsection{Distribution Validation}
\begin{figure*}[!t]
\centering
\includegraphics[scale=0.25]{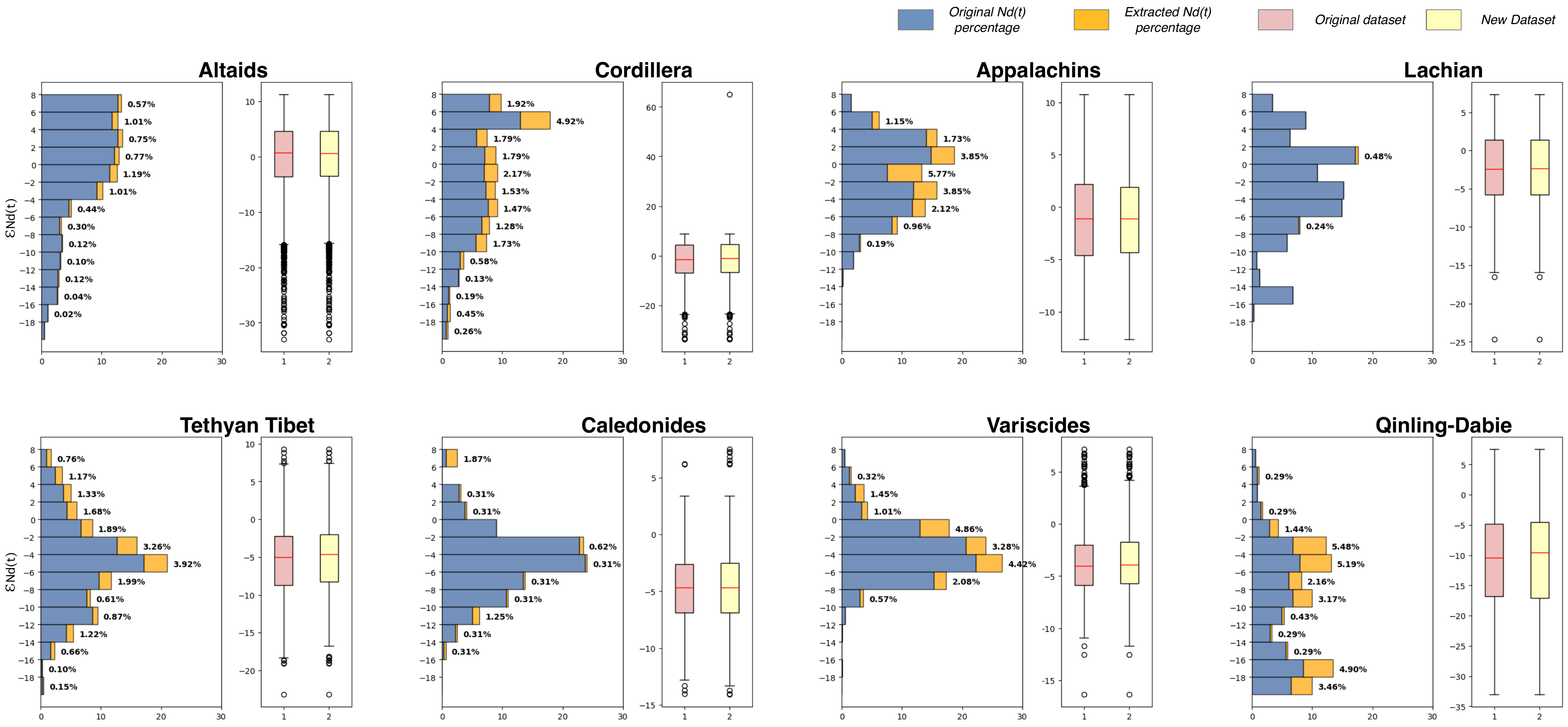}
\caption{Distribution validation of the original dataset and the dataset obtained through our methodology, illustrating the frequency distribution of $\epsilon$ Nd(t) values for the eight orogens. The blue diagram depicts the $\epsilon$ Nd(t) percentages from the original dataset, while the orange diagram reflects the percentages from our acquired dataset. In the box plot representation, the pink box delineates the $\epsilon$ Nd(t)value distribution of the original dataset, whereas the yellow box characterizes the $\epsilon$ Nd(t) value distribution from our newly acquired dataset.}
\label{isotope}
\end{figure*}

In Figures~\ref{isotope}, we present a distribution analysis of sample distributions within isotopic domains ($\epsilon$Nd (t)) contrasted with two-stage Nd-depleted mantle model ages (TDM2). The robustness of our comparative analysis is confirmed by a pronounced linear correlation across eight global orogens, as demonstrated by Wang et al. (2023) \cite{wang2023quantitative}. Using a refined and methodologically advanced approach, we have increased the granularity of the data representation, with pronounced improvements observed in geologically complex regions, including the Cordillera, Appalachians, Tethyan Tibet, Variscides, and Qinling-Dabie orogens. Empirical analysis shows an average data improvement of 22.45\%, a significant increase that adds resolution and depth to the interpretability of the dataset.

\subsection{Efficiency Evaluation}

\begin{figure*}[!t]
\centering
\includegraphics[scale=0.3]{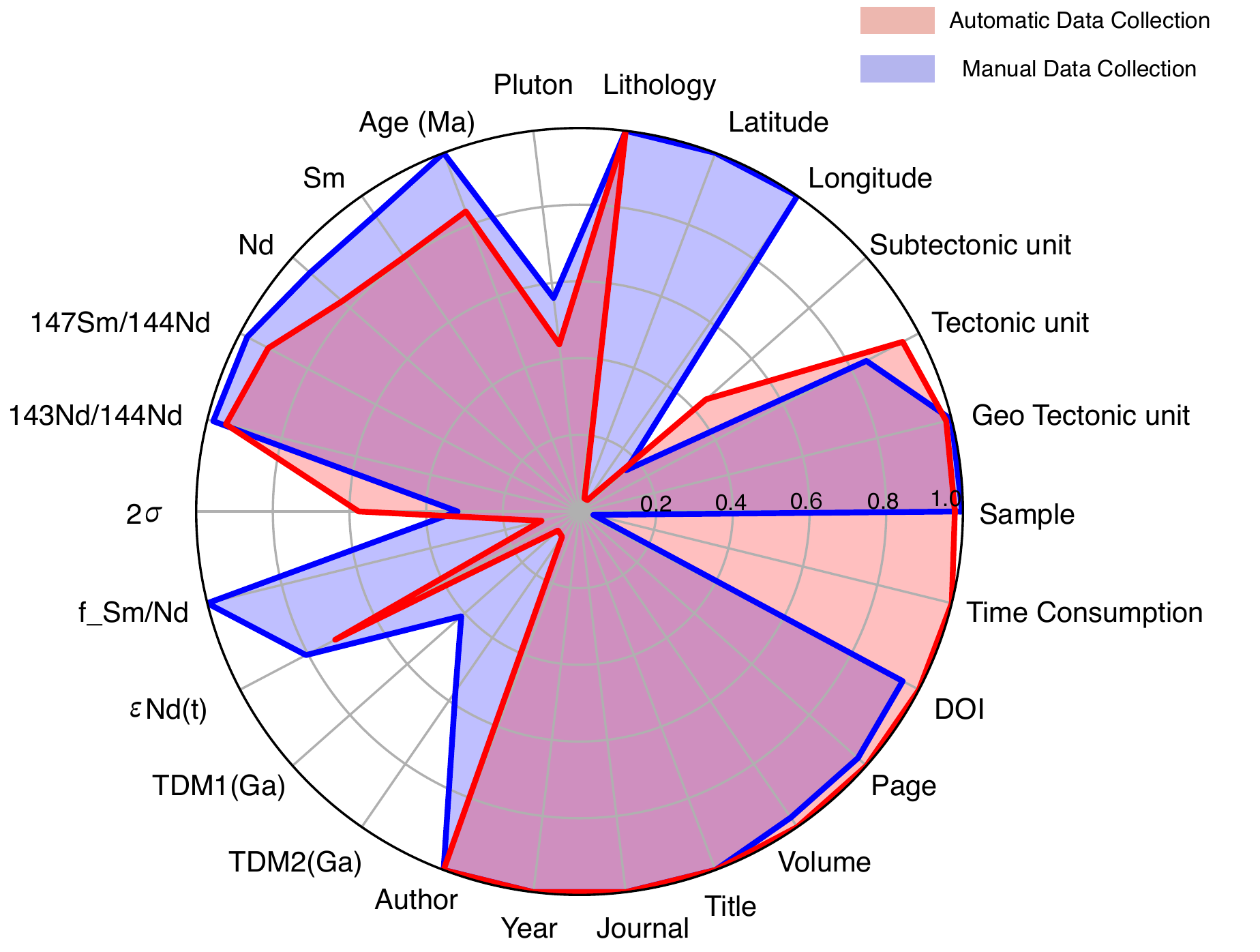}
\caption{Comparison of data filling rate and time consumption between manual compilation and our automatic data collection method.}
\label{statistics}
\end{figure*}

As shown in Figure~\ref{statistics}, our research delved into an intensive analytical evaluation focused on unraveling the competence of various data collection modalities. Our research juxtaposed automatic and manual data collection techniques and anchored our evaluation on key variables, including time efficiency and data fill rate. This included vital components such as position, attribute, age, sample, and meta-information. Our in-depth comparative analysis revealed marked differences between the two methodologies. One notable observation was the fragmentation of critical data, particularly age and position, interspersed among the various textual and graphical components embedded in the scientific literature. The diverse nature of this information dissemination poses a notable challenge to automatic data extraction, resulting in noticeable variances compared to the precision and coherence inherent in manual extraction techniques.

However, the extraction of detailed specimen information, mainly associated with sample information, such as tectonic units, is one domain where the precision and speed of automatic mechanisms are strikingly evident. Incorporating an elaborate, carefully curated keyword dictionary enhances the adeptness of automatic extraction, making it faster and significantly more comprehensive. Our focused examination of the temporal dynamics inherent in data extraction and assimilation accentuates an empirical delineation of this enhanced efficiency. To contextualize the time efficiency attributable to our automatic approach, we conducted a comparative analysis, drawing on a dataset comprising 9,000 manually collected and curated Sm-Nd-related records. These records were gleaned utilizing a similar keyword schema as employed in the automatic approach. The findings were illuminating – the manual process was 27 times more time-intensive. This distinction accentuates the functional efficacy of automatic processes, particularly in contexts where the alacrity of data extraction and the integrality of information are paramount. The detailed comparison of data fill rates is illustrated in Table~\ref{filingrate}.

\subsection{Limitations}
Throughout our automatic data collection process, we encountered several challenges that called into question the reliability of the automatic data collection process. A predominant challenge was encoding older literature into PDFs via image scanning. This encoding method introduced complexities to data retrieval using traditional PDF parsing techniques. The traditional paradigm for data extraction was firmly rooted in using Optical Character Recognition (OCR) tools. However, the nuances of specific symbols and specialized terms in academic literature tables often led to data misinterpretation by general-purpose OCR tools, compromising the accuracy of data capture.

The landscape of table data definitions presented challenges primarily due to an absence of standardization. This lack necessitated the compilation of an array of header definitions, which, in turn, introduced variability, constraining the efficacy of our automatic table data extraction mechanism from academic literature. Within specific academic disciplines, formulating a targeted keyword list became pivotal, enhancing the efficiency of table data acquisition.

\begin{table*}[t]
\begin{center}
\begin{tabular}{lcccccc}
\toprule
Kewords     & Manual Collection  & Our Method  \\ \hline
Sample  & \textbf{0.998} & 0.979    \\
GeoTectonic unit      &  \textbf{1.000}     &   0.985    \\
Tectonic unit       &0.845&\textbf{0.952}\\
Subtectonic unit       &0.162&\textbf{0.441}\\
Longitude       &   \textbf{1.000}    &0.036\\
Latitude&\textbf{1.000}&0.036\\
Lithology&\textbf{1.000}&\textbf{1.000}\\
Pluton&\textbf{0.561}&0.439\\
Age (Ma)&\textbf{1.000}&0.837\\
Sm&\textbf{0.939}&0.806\\
Nd&\textbf{0.939}&0.827\\
${ }^{147} \mathrm{Sm} /{ }^{144} \mathrm{Nd}$&\textbf{0.979}&0.917\\
${ }^{143} \mathrm{Nd} /{ }^{144} \mathrm{Nd}$&\textbf{0.984}&0.950\\
2$\sigma$&0.318&\textbf{0.576}\\
$f_{Sm/Nd}$&\textbf{1.000}&0.102\\
$\epsilon$Nd(t)&\textbf{0.806}&0.720\\
TDM1 (Ga)&\textbf{0.412}&0.075\\
TDM2 (Ga)&\textbf{0.570}&0.080\\
Author&\textbf{1.000}&\textbf{1.000}\\
Year&\textbf{1.000}&\textbf{1.000}\\
Journal&\textbf{1.000}&\textbf{1.000}\\
Title&\textbf{1.000}&\textbf{1.000}\\
Volume&0.970&\textbf{1.000}\\
Page&0.970&\textbf{1.000}\\
DOI&0.952&\textbf{1.000}\\ \hline
Average&\textbf{0.856}&0.710\\
\bottomrule
\end{tabular}
\end{center}
\caption{Comparative analysis of the filling rate between manual collection and our automatic method across a dataset of 9,000 samples. The best performance are \textbf{bold} within each comparison.}\protect
\label{filingrate}
\end{table*}

\section{Usage Notes}
The Sm-Nd isotope dataset provides a valuable resource for studies of orogens, their compositional architecture, and crustal growth patterns \cite{goldstein1984sm,blanchet2019database,condie2013preservation,cawood2009accretionary,cawood2016linking,collins2011two,collins2002hot}. This dataset facilitates a nuanced understanding of orogen characterization and categorization \cite{dickin1998nd,dickin2000crustal, wang2009nd}. For example, the Nd isotope mapping method, which uses Sm-Nd data in conjunction with age and position information, provides a robust approach to studying crustal growth \cite{gruau1991origin,lambert1994re}. \cite{wang2023voluminous, wang2023quantitative} presented Nd isotope mapping results from eight emblematic Phanerozoic orogens. Their research delineates the crustal growth patterns inherent to each orogenic belt and introduces an innovative methodology that quantitatively characterizes orogens based on their compositional architecture via isotopic mapping. This method elucidates the intricate relationships between orogenesis and continental growth. In addition, we have pioneered our automated tabular data collection method, which significantly increases the efficiency of data collection and results in a richer, more comprehensive data set.

\section{Code Availability}

The tabular collection code is available from \url{https://github.com/sjtugzx/tabularDataCollection}. Instructions for use are available with the code.

\bibliography{sample}

\section{Author contributions statement}
Zhixin Guo: Conceptualization (equal); methodology (equal); software (equal); validation (lead); writing - original draft (lead). Tao Wang: Conceptualization (equal); data validation (equal). Chaoyang Wang: Conceptualization (equal); data validation (equal). Jianping Zhou: Software (equal); writing - review and editing (equal). Guanjie Zheng: Conceptualization (equal); formal analysis (equal); methodology (equal); project administration (equal); supervision (lead); writing - review and editing (equal). Xinbing Wang: Project administration (equal); supervision (equal). Chenghu Zhou: Project administration (equal).

\section{Competing interests}
The authors declare no competing interests.

\end{document}